# Focusing inversion techniques applied to electrical resistance tomography in an experimental tank

G. Pagliara[1], G. Vignoli[2]

1 Math4Tech, Mathematics Department, University of Ferrara, Italy
2 L.A.R.A. Group, Earth Sciences Department, University of Ferrara, Italy
Corresponding author: giulio.vignoli@unife.it

ABSTRACT : We present an algorithm for focusing inversion of electrical resistivity tomography (ERT) data. ERT is a typical example of ill-posed problem. Regularization is the most common way to face this kind of problems; it basically consists in using a priori information about targets to reduce the ambiguity and the instability of the solution. By using the minimum gradient support (MGS) stabilizing functional, we introduce the following geometrical prior information in the reconstruction process: anomalies have sharp boundaries.

The presented work is embedded in a project (L.A.R.A.) which aims at the estimation of hydrogeological properties from geophysical investigations. L.A.R.A. facilities include a simulation tank (4 m x 8 m x 1.35 m); 160 electrodes are located all around the tank and used for 3-D ERT. Because of the large number of electrodes and their dimensions, it is important to model their effect in order to correctly evaluate the electrical system response. The forward modelling in the presented algorithm is based on the so-called complete electrode model that takes into account the presence of the electrodes and their contact impedances.

In this paper, we compare the results obtained with different regularizing functionals applied on a synthetic model.

## 1. Introduction

ERT is a very widely used technique: its applications go from mining exploration to detection and mapping of subsurface contaminant plumes. To determine the resistivity distribution from measurements of potential differences is a non-unique problem and its numerical solution is unstable: small variations in the data can cause large variations in the solution. Commonly used inversion methods provide unique and stable solutions by introducing the appropriate stabilizing functional (stabilizer). The main aim of the stabilizer is to incorporate a priori knowledge in the inversion process. Over the last decade several different stabilizers have been introduced (Zhdanov, 2002). These new stabilizers permit reconstruction of blocky structures with abrupt change of properties. They generate clearer and more focused images of the anomalies than the conventional maximum smoothness functionals. For example, it was shown that the minimum support (MS) functional can be very useful in the solution of different geophysical inverse problems (Portniaguine and Zhdanov, 1999; Vignoli and Zhdanov, 2004; Vignoli and Zanzi, 2005). This particular functional selects the desired stable solution with the following characteristic: anomalies have sharp boundaries. In this work, we apply a special version of MS stabilizer, the MGS functional, to ERT on a finite volume (Blaschek, 2005). Moreover we compare MGS solution with the results provided, respectively, by the more traditional minimum norm (MN) and the total variation (TV) stabilizers.

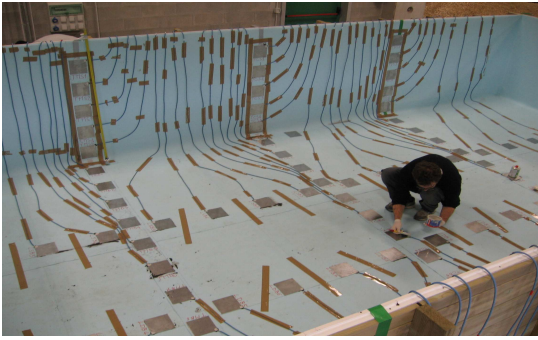 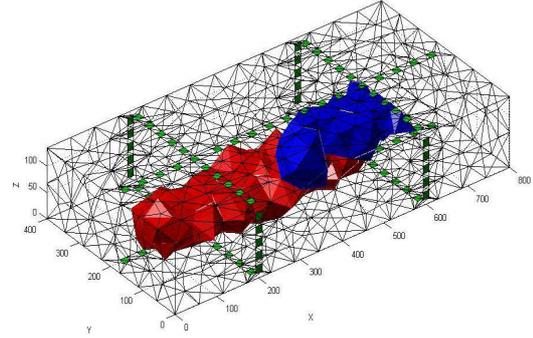

Fig. 1. Set up of one of the simulation tanks at L.A.R.A. laboratory. It is equipped with 160 electrodes.

Fig. 2. Synthetic model of our numerical test; it consists of two anomalies (red and blue bodies); data are measured by 96 plane electrodes (green).

We simulate to collect our data over a finite volume (a tank, as in Fig. 1, representing L.A.R.A. facility) surrounded by an insulator medium (the tank lining and the air). In our synthetic model (Fig. 2), we imagine to deal with a large number of electrodes (96) with significant dimensions (0.15 m x 0.15 m). For all this reasons, we have to model also the electrodes effect. The forward modelling in the presented algorithm is based on the so-called complete electrode model that takes into account the presence of the electrodes and their contact impedances. The advantages of this model are that it accounts for the fact that current is applied through electrodes, which are discrete and are high conductive shortcuts for the current; moreover, it does not ignore the effect of the contact impedance and the fact that injected current density is not constant at all at electrode-object interfaces.

## 2. ERT forward and inverse problem
### 2.1. Complete electrodes model

ERT consists of using electrodes (metal stakes knocked into the earth or, as in our case, plates attached on the surface of an object) to inject direct electric current into the ground and measure the corresponding electrical potential at the surface. The measured data depend on the spatial distribution of electrical conductivity (the parameter). The process which maps the parameter to the measured data (the forward mapping) depends nonlinearly on the conductivity distribution.

To model the physical system, we made the following assumptions: 1) the investigated object is linear (i.e., the physical properties are independent on applied field strength); 2) the medium is microscopically isotropic; 3) electrodes are perfect conductors on the surface of the investigated object, which is embedded in an insulator medium. We have adopted the complete electrode model (Polydorides, 2002): on one hand, it takes into account that the actual measured potential at $l^{th}$ electrode is equal to the sum of the potential on the boundary surface underneath the electrode and the potential drop across the electrode contact resistance[*]; on the other hand, it models the shunting effect of the electrodes, i.e., the metal electrodes themselves provide a low resistance path for current (Fig. 3a). Moreover, differently from what it is usually assumed, the current density is far from being constant underneath electrodes (Fig. 3); thus, instead of imposing a condition on the current density, we use a weaker condition on the total injected current.

---
[*] Contact resistance is due to a common electrochemical effect that may take place at the contact between the electrode and the object. It consists in the formation of a thin resistive layer between the electrode and the object; it vanishes when the electrode is in touch with an ohmic conductor.

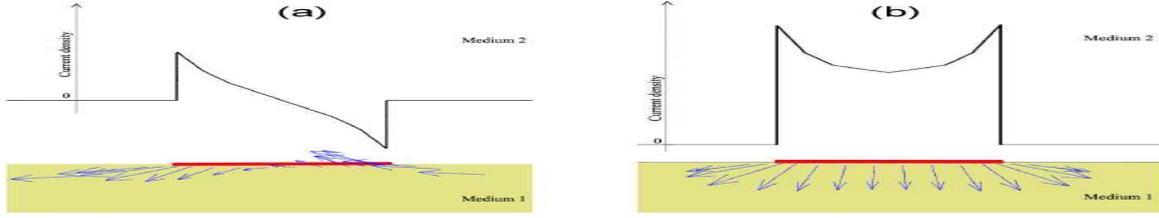

Fig. 3. Current density at a measuring (a) and injecting (b) electrode. On the top, a qualitative plot concerning the current density behaviour along an electrode (in red). On the bottom, the current density vectors (blue arrows) are shown in proximity to the same electrode.

### 2.2. Regularized inversion: three possible choices for the stabilizer

A common way to solve inverse problems is by minimization of the Tikhonov parametric functional of the model parameter vector $m : P^\alpha(m) = \phi(m) + \alpha s(m)$; it combines least square data misfit between the calculated data $A(m)$ and the observed data $d$ : $\phi(m) = \|A(m) - d\|_{L_2}^2$, and the stabilizer $s(m)$, whose function is to select a correctness subset from the space of all possible models. There are several different choices for the stabilizer and, of course, different stabilizers produce different solutions. In this paper, we analyse: 1) the MN stabilizer, which is proportional to the difference between the model $m$ and an appropriate a priori model $m_{apr}$ : $s_{MN}(m) = \int_\Omega |m(r) - m_{apr}(r)|^2 dr$ ; 2) the TV stabilizer: $s_{TV}(m) = \int_\Omega |\nabla(m(r) - m_{apr}(r))| dr$, commonly used in order to recover non-smooth targets; 3) the MGS stabilizer, which is equal to the volume (support) where the variation of the difference between the current model $m$ and the a priori model $m_{apr}$ is non-zero: $s_{MS}(m) = \int_\Omega \frac{(\nabla(m(r) - m_{apr}(r)))^2}{(\nabla(m(r) - m_{apr}(r)))^2 + e^2} dr$, with e focusing parameter. It is easy to demonstrate that $s_{MGS} \to \text{support}\{\nabla(m - m_{apr})\}$ if $e \to 0$.

The regularization parameter $\alpha$ in $P^\alpha$ describes a trade-off between the best fitting and the most reasonable stabilization. To find the optimal $\alpha$, given a set of values $\{\alpha_k\}$, we compute, for each $\alpha_k$, the corresponding model $m_{\alpha_k}$ and misfit $\phi(m_{\alpha_k})$ minimizing $P^{\alpha_k}(m)$. The optimal value is the number $\alpha_{k_0}$ for which the equation $\phi(m_{\alpha_{k_0}}) = \delta$ is satisfied, with $\delta$ being the noise level in the observed data.

## 3. Comparison of different stabilizer inversion

Let us compare inversion results obtained using the three stabilizers discussed overleaf (Fig. 5, 6, 7). We imagine to "collect" synthetic data using the electrode configuration depicted in Fig. 2. Electrodes are imagined to be placed on the surface of L.A.R.A. simulation tank. The true model (Fig. 4) consists of two anomalies: respectively 20% more (the red one) and less (the blue one) resistive than background. Detecting them is particularly difficult because they partially shield each other. MGS generates the best result: its reconstruction is sharper than TV solution; it can recover properly even small features (e.g., see slices at $z \cong 62 \text{ cm}, 72 \text{ cm}, 82 \text{ cm}$ in Fig, 7 and compare them with the corresponding planes in Fig. 5 and 6.); besides, MGS provides a quite reliable estimation of the correct resistivity value.

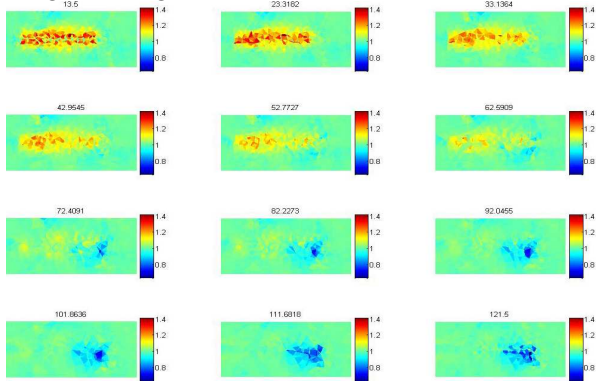
Fig. 4. Longitudinal slices of the true model.

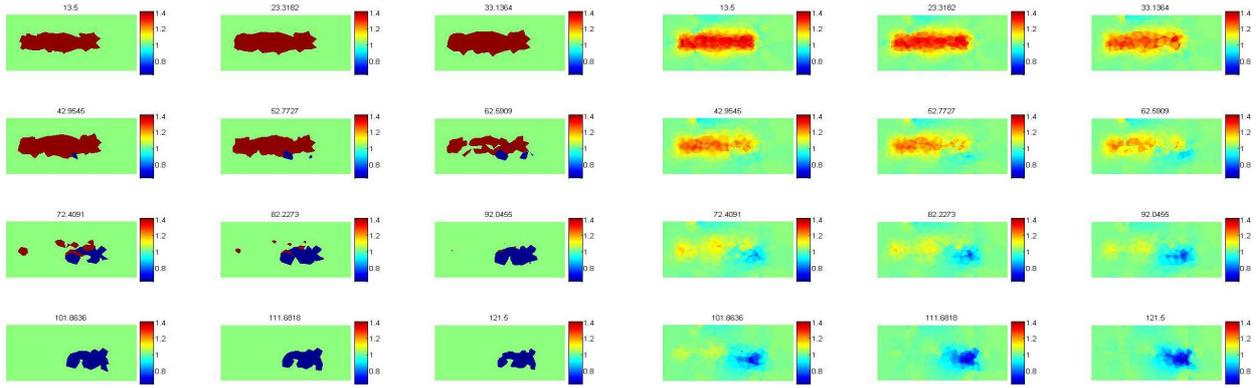
Fig. 6. Longitudinal slices of TV result.

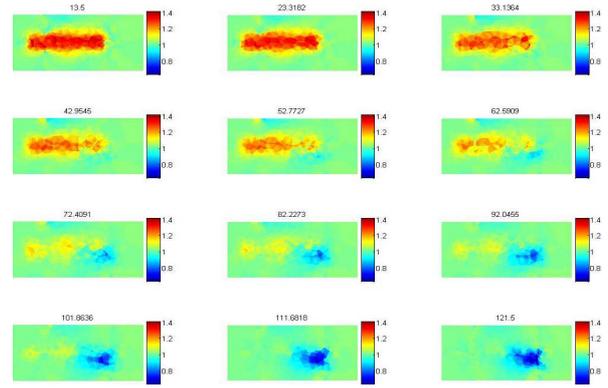
Fig. 5. Longitudinal slices of MN result.

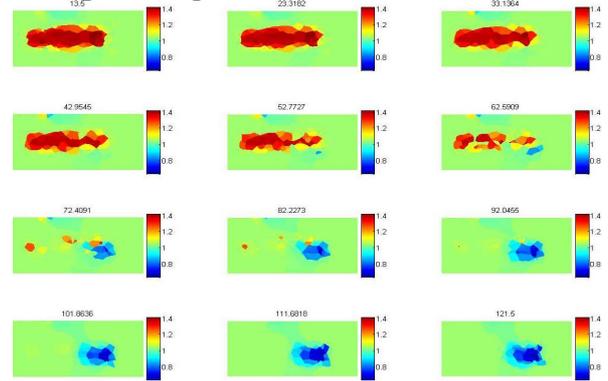
Fig. 7. Longitudinal slices of MGS result.

## 4. Conclusion

Stabilization of inversion methods with smoothing functionals can result in oversmoothed reconstruction of object properties (Fig. 5). We suggest the use of the $s_{MGS}$ or $s_{TV}$ to preserve sharp features in inverted models. However, it is evident that using MGS provides even better images of blocky targets when prior information about the anomalies is available; the example shows that this stabilizer can also identify the right resistivity values.

The importance of complete electrode model sophistication will be evaluated when real data of L.A.R.A. facility will be available. In any case, because of the large number of electrodes and their dimensions, it seems very important not to neglect their influence.